\documentclass[english,10pt,aps,pra,twocolumn,floatfix]{revtex4-1}

\usepackage[T1]{fontenc}
\usepackage{amsmath}
\usepackage{amssymb}
\usepackage{amsfonts}
\usepackage{bm}
\usepackage{bbm}
\usepackage{epsfig}
\usepackage{grffile}
\usepackage{graphics}

\usepackage[usenames,dvipsnames]{color}
\definecolor{dblue}{rgb}{0,0.1,.6}

\usepackage[colorlinks=true,citecolor=dblue,linkcolor=dblue,urlcolor=dblue]{hyperref}
\usepackage[all]{hypcap}

\newcommand{\ud}{\mathrm{d}}
\newcommand{\id}{\mathbbm{1}}

\newcommand{\bra}{\langle}
\newcommand{\ket}{\rangle}
\newcommand{\mc}[1]{\mathcal{#1}}

\newcommand{\hH}{\hat{H}}
\newcommand{\hS}{\hat{S}}
\newcommand{\hvS}{\hat{\vec{S}}}

\newcommand{\hA}{\hat{A}}
\newcommand{\hB}{\hat{B}}
\newcommand{\hF}{\hat{F}}
\newcommand{\veps}{\varepsilon}

\renewcommand{\vec}[1]{{\boldsymbol{#1}}}
\newcommand{\s}{\sigma}
\newcommand{\bs}{\bar{\sigma}}
\newcommand{\vs}{\vec{\sigma}}

\newcommand{\pred}{\text{pred}}

\newcommand{\trunc}{\text{trunc}}
\newcommand{\SU}{\mathrm{SU}}

\newcommand{\duke} {Department of Physics, Duke University, Durham, North Carolina 27708, USA}

\newcommand{\Title} {Infinite boundary conditions for response functions and limit cycles in iDMRG, demonstrated for bilinear-biquadratic spin-1 chains}
\newcommand{\Authors}
{
\author{Moritz Binder}
\affiliation{\duke}
\author{Thomas Barthel}
\affiliation{\duke}
}
\newcommand{\Date} {May 25, 2018}

\begin{document}

\title{\Title\vspace{-0.5em}}
\Authors

\begin{abstract}
Response functions $\bra A_x(t) B_y(0)\ket$ for one-dimensional strongly correlated quantum many-body systems can be computed with matrix product state (MPS) techniques. Especially, when one is interested in spectral functions or dynamic structure factors of translation-invariant systems, the response for some range $|x-y|<\ell$ is needed.
We demonstrate how the number of required time-evolution runs can be reduced substantially: (a) If finite-system simulations are employed, the number of time-evolution runs can be reduced from $\ell$ to $2\sqrt{\ell}$.
(b) To go beyond, one can employ infinite MPS (iMPS) such that two evolution runs suffice. To this purpose, iMPS that are heterogeneous only around the causal cone of the perturbation are evolved in time, i.e., the simulation is done with infinite boundary conditions. Computing overlaps of these states, spatially shifted relative to each other, yields the response functions for all distances $|x-y|$.
As a specific application, we compute the dynamic structure factor for ground states of bilinear-biquadratic spin-1 chains with very high resolution and explain the underlying low-energy physics.
To determine the initial uniform iMPS for such simulations, infinite-system density-matrix renormalization group (iDMRG) can be employed. We discuss that, depending on the system and chosen bond dimension, iDMRG with a cell size $n_c$ may converge to a non-trivial limit cycle of length $m$. This then corresponds to an iMPS with an enlarged unit cell of size $m n_c$.
\end{abstract}

\date{\Date}
\maketitle

\section{Introduction}\label{sec:intro}\vspace{-0.5em}
Spectral functions and dynamic structure factors give insights into the many-body physics of strongly correlated quantum systems by providing detailed information about the low-lying excitations. Experimentally, these quantities can be measured directly, for example, by using inelastic neutron-scattering or ARPES techniques. Their calculation with numerical methods remains challenging. For one-dimensional (1d) systems, density-matrix renormalization group (DMRG) algorithms allow us to efficiently determine precise approximations of ground-state wavefunctions, either as matrix product states (MPS) for finite systems or as infinite MPS (iMPS) in the thermodynamic limit \cite{White1992-11,White1993-10,McCulloch2008_04}. In combination with time-evolution algorithms \cite{Vidal2003-10,White2004,Daley2004}, one can compute real-time response functions, from which spectral functions can be obtained by Fourier transform. One challenge in this approach is the growth of entanglement during time evolution, which leads to a corresponding
growth of the computation cost with time. A simple trick allows to double the maximum reachable times, but requires a separate simulation for each distance for which the response function is needed.

Here, we introduce a new scheme for finite-system simulations of translation-invariant models that reduces the required number of time-evolution runs significantly. In addition, we show how iMPS simulations can be used to reduce the number of required runs to only two. If we apply a local perturbation to an initial uniform iMPS and evolve the state in time, Lieb-Robinson bounds \cite{Lieb1972-28} guarantee that the perturbation only has a significant effect on sites within a causal cone, a finite spatial region that grows linearly with time. We can hence simulate the time evolution using a finite heterogeneous window with infinite boundary conditions, a technique that has been used before to study quantum quenches \cite{Phien2012-86,Phien2013-88}. The difference in our approach is that we shift the heterogeneous windows of two iMPS relative to each other to evaluate the response functions for all required distances between perturbation and measurement. We employ this new technique to compute high-resolution dynamic spin structure factors for bilinear-biquadratic spin-1 chains. We explain the corresponding low-energy physics. Note that after the main parts of this work were completed, Ref.~\cite{Lange2018-97} appeared, which introduces a similar technique for the simulation of finite-temperature response functions with purifications.

Different algorithms exist to obtain an approximation of the ground state in form of a uniform iMPS. We use infinite-system DMRG (iDMRG) \cite{McCulloch2008_04} which is rather efficient and straightforward to implement. Alternatively, one can employ the uniform variational iMPS algorithm of Ref.~\cite{Zauner2018-97} or imaginary-time evolution using the infinite time-evolving block decimation algorithm \cite{Vidal2007-98,Orus2008-78}. As the ground-state calculation is followed by a real-time evolution which typically dominates the computation cost, possible differences in convergence speeds for the ground-state computation are not so important.

We address the convergence of the iDMRG algorithm for different phases of the bilinear-biquadratic spin-1 model and sizes of the iDMRG unit cell $n_c$. We observe that under certain conditions, the algorithm fails to converge to a simple fixed point, but can rather converge to a nontrivial limit cycle of length $m$. Based on our analysis of corresponding fidelities, this is to be interpreted as the proximity to a low-entangled iMPS solution with a larger unit cell of size $mn_c$. We also discuss how the use of symmetric initial states for iDMRG can result in substantially increased bond dimensions and degeneracies of the MPS transfer operator that can cause problems in the wavefunction orthogonalization.

The structure of this work is as follows. In Sec.~\ref{sec:bilinbiquad}, we briefly summarize the different phases of the bilinear-biquadratic spin-1 chain. Section~\ref{sec:response} reviews the calculation of response functions for translation-invariant systems in the linear response regime using matrix product state methods. In Sec.~\ref{sec:finite-system}, we describe schemes to calculate response functions with finite-system simulations. In Sec.~\ref{sec:infinite-system}, we introduce an efficient scheme using infinite boundary conditions that reduces the number of required time-evolution runs. We apply this technique to the calculation of dynamic spin structure factors for the bilinear-biquadratic spin-1 chain in Sec.~\ref{sec:results} and interpret the different features. In Sec.~\ref{sec:cycles}, we discuss nontrivial limit cycles in the iDMRG algorithm. We summarize and conclude in Sec.~\ref{sec:discussion}.

\section{Bilinear-biquadratic spin-1 chain}\label{sec:bilinbiquad}\vspace{-0.8em}
As a specific application, we simulate $\SU(2)$ symmetric bilinear-biquadratic spin-1 chains
\begin{equation}\label{eq:Hblbq}\textstyle
	\hH=\sum_{i}\big[\cos\theta\,\hvS_i\cdot\hvS_{i+1}+\sin\theta\,(\hvS_i\cdot\hvS_{i+1})^2\big].
\end{equation}
For $\pi/2<\theta< 5\pi/4$, the system is in a gapless ferromagnetic and, for $5\pi/4<\theta<7\pi/4$, in a gapped dimerized phase \cite{Barber1989-40,Kluemper1989-9,Xian1993-5,Laeuchli2006-74}. For $-\pi/4<\theta< \pi/4$, it is in the gapped Haldane phase \cite{Haldane1983-93} which includes the AKLT point $\tan\theta=1/3$ where the ground state is an MPS with bond dimension $D=2$ \cite{Affleck1987-59}. The Haldane phase is an example for symmetry-protected topological order \cite{Gu2009-80,Pollmann2012-85}. For $\pi/4\leq\theta\leq \pi/2$, the system is in a gapless spin quadrupolar phase \cite{Fath1991-44,Fath1993-47,Itoi1997-55,Laeuchli2006-74}, and, in the vicinity of the integrable point $\theta=\pi/4$, the long-range physics is governed by the level-1 $\SU(3)$ Wess-Zumino-Witten model with marginally irrelevant perturbations \cite{Itoi1997-55,Witten1984-92}. In this critical phase, period-three spin quadrupolar correlations dominate. In particular, approaching the gapless phase from $\theta=0$, the single-magnon excitation with the minimum (Haldane) gap at $k=\pi$ deforms with increasing $\theta$. The minimum switches to $k=2\pi/3$ and the gap at this momentum closes at the critical point $\theta=\pi/4$.

The model \eqref{eq:Hblbq} describes several quasi-1d quantum magnets. Some examples are
CsNiCl$_3$ \cite{Tun1990-42,Zaliznyak2001-87} and Ni(C$_2$H$_8$N$_2$)$_2$NO$_2$ClO$_4$ (NENP) \cite{Ma1992-69,Regnault1994-50}
which are to a good approximation Heisenberg antiferromagnets ($\theta=0$) or LiVGe$_2$O$_6$ \cite{Millet1999-83,Lou2000-85} which features a sizable biquadratic coupling.

\section{Response functions for translation-invariant systems}\label{sec:response}\vspace{-0.8em}
The response of a system to a time-dependent perturbation $\hH\to\hH'(t)=\hH+f(t)\hB_y$ around site $y$ can be characterized by its influence on the expectation value of an observable $\hA_x$, located around site $x$. According to linear response theory, the effect of weak perturbations is determined by time-dependent correlation functions,
\begin{equation*}
	\bra\hA_x\ket_t=\bra\hA_x\ket_{0} -i\int_{-\infty}^t\hspace{-2ex}\ud t'\,f(t')\bra[\hA_x(t),\hB_y(t')]\ket_0+\mc{O}(f^2)
\end{equation*}
which is the fluctuation-dissipation theorem. Here, $\bra\dots\ket_{0}$ denotes the expectation value with respect to an initial $\hH$-equilibrium state and the right-hand side uses the Heisenberg picture.
In this work, we consider systems at zero temperature, i.e., response functions have the form $\bra\psi|\hA_x(t)\hB_y(t')|\psi\ket$ if the ground state $|\psi\ket$ is non-degenerate, and are given by a corresponding average in the case of degeneracies.

In translation-invariant autonomous systems, the response functions only depend on $x-y$ and $t-t'$. They can hence be parametrized, equivalently, by momentum $k$ and frequency $\omega$, i.e., the Fourier transform with respect to space and time
\begin{gather}\label{eq:Skw}
	S(k,\omega) = \sum_{x}e^{-ikx}\int\ud t\, e^{i\omega t}S(x,t)\\\nonumber
	\text{with}\quad S(x,t) = \bra\psi|\hA_x(t)\hB_0(0)|\psi\ket.
\end{gather}
These are typical experimental observables, e.g., measured in ARPES and neutron scattering experiments.

For 1d systems, ground states $|\psi\ket$ can be determined efficiently in MPS form using the DMRG method. The simplest way to compute response functions would be to apply $\hB_0$ to $|\psi\ket$ and use time-dependent DMRG (tDMRG) to obtain an MPS approximation of $|\psi_{\hB_0}(t)\ket:=e^{-i\hH t}\hB_0|\psi\ket$. Then, the response function is obtained by evaluating matrix elements
\begin{equation}\label{eq:Ssimple}
	S(x,t)=e^{iE_0t}\bra\psi|\hA_x|\psi_{\hB_0}(t)\ket
\end{equation}
with the ground-state energy $E_0$.
One issue with this approach is that, in states $|\psi_{\hB_0}(t)\ket$, correlations spread in a causal cone emanating from $(x,t)=(0,0)$. As a consequence, entanglement entropies for bipartitions that cut this region grow and tDMRG computation costs increase accordingly, limiting the maximum reachable times.

There is a simple trick to increase (typically double) the accessible time range. By also computing $|\psi_{\hA^\dag_x}(-t)\ket$, we have 
\begin{equation}\label{eq:Sdouble}
	S(x,t_1+t_2)=e^{iE_0(t_1+t_2)}\bra\psi_{\hA^\dag_x}(-t_1)|\psi_{\hB_0}(t_2)\ket.
\end{equation}
While this is very advantageous concerning the reachable times, there is one drawback compared to Eq.~\eqref{eq:Ssimple}. Often, one needs $S(x,t)$ for some range of distances $x$. The typical situation is that we want to compute spectral functions $S(k,\omega)$ and hence need all $S(x,t)$ that are not negligible for the spatial Fourier transform in Eq.~\eqref{eq:Skw}, corresponding to a time-dependent range $|x|<\ell$. While for Eq.~\eqref{eq:Ssimple}, only a single time-evolution run is needed, assuming reflection symmetry, we need $\sim \ell$ runs for Eq.~\eqref{eq:Sdouble}. We will show in the following how this can be avoided.

Note that the trick of Eq.~\eqref{eq:Sdouble}, in a generalized form, also works for non-zero temperatures \cite{Barthel2012_12,Barthel2013-15,Binder2015-92}. Another device to reach longer times, which is by now a standard tool, is to extrapolate the simulation data through linear prediction \cite{White2008-77,Barthel2009-79b}.

\section{Finite-size simulations}\label{sec:finite-system}
\begin{figure}[t]
\includegraphics[width=\columnwidth]{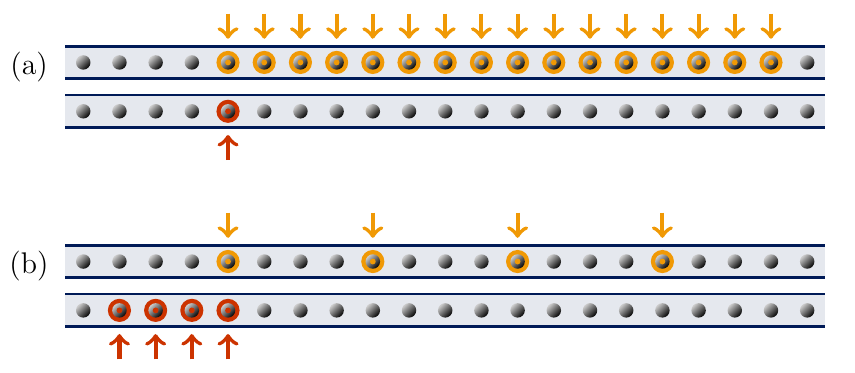}
\caption{\label{fig:finite-size_schemes}\textbf{Finite-size schemes.} (a) In finite-size simulations, response functions $S(x,t)$ for a certain spatial range $0\leq x<\ell$ can be evaluated according to Eq.~\eqref{eq:Sdouble} by evolving $\hB_0|\psi\ket$ and $\hA^\dag_x|\psi\ket$, and computing their overlaps. This requires $\ell+1$ evolution runs. (b) A more efficient method is to evolve $\hB_{x'}|\psi\ket$ and $\hA^\dag_{x''}|\psi\ket$ for some strategically chosen sites $x'$ and $x''$ as in Eq.~\eqref{eq:SdoubleFS}. This ``square-grid scheme'' reduces the number of required evolution runs to $\approx 2\sqrt{\ell}$.}
\end{figure}
\subsection{Reduced number of evolution runs in the square-grid scheme}
The number of time-evolution runs, required to compute $S(x,t)$ for $|x|<\ell$ with maximum times as in Eq.~\eqref{eq:Sdouble}, can be reduced from $\sim\ell$ to $\sim 2\sqrt{\ell}$. For simplicity, reflection symmetry is assumed such that $x\geq 0$ is sufficient.
Let $n=\operatorname{ceil}(\sqrt{\ell})$. Do $n$ time-evolution runs to compute $|\psi_{\hB_j}(t_2)\ket$ for $j=0,-1,\dotsc,-(n-1)$. Then, do $n$ runs to compute $|\psi_{\hA^\dag_{in}}(-t_1)\ket$ for $i=0,1,\dotsc,n-1$ as indicated in Fig.~\ref{fig:finite-size_schemes}. Their mutual overlaps yield
\begin{equation}\label{eq:SdoubleFS}
	S(x=in-j,t_1+t_2)=e^{iE_0(t_1+t_2)}\bra\psi_{\hA^\dag_{in}}(-t_1)|\psi_{\hB_j}(t_2)\ket.
\end{equation}
for $x=0,\dotsc,n^2-1$. To avoid finite-size effects, sites $i$ and $j$ need to be at sufficient distance from the boundaries in a region where the state is approximately translation invariant. Finite-size effects can be reduced further by employing smooth boundary conditions \cite{Vekic1993-71}.
The described ``square-grid scheme'' can be easily adapted for states $\psi$ with a unit cell of size $n_c>1$ in the bulk.

\subsection{Fourier-transformed operators}
Sometimes, one is not interested in the full momentum dependence of the response function but only, say, a particular momentum $k$ or a few momenta. In such cases, an alternative to computing the response $S(x,t)$ and Fourier transform is to directly determine the Fourier transformed correlator \cite{Barthel2013-15}
\begin{equation}
	\bra\psi|\hA_k(t)\hB_0(0)|\psi\ket\quad\text{with}\quad
	\hA_k=\sum_{x}e^{-ikx}\hA_x.
\end{equation}
If $\hA_x$ is a single-site operator, $\hA_k$ can be written as a matrix product operator (MPO) with bond dimension $D=2$. For a single momentum $k$, we can use the analog of Eq.~\eqref{eq:Sdouble} to compute $S(k,t)$ with only two evolution runs. This approach has one drawback. While the entanglement growth in states $|\psi_{\hA^\dag_x}(-t)\ket$ is restricted to a causal cone centered at site $x$, entanglement entropies typically grow for all bonds in states $|\psi_{\hA^\dag_k}(-t)\ket$. In comparison, the computation costs are hence typically increased by a factor proportional to the system size or to $\ell$ if the sum in the definition of $\hA_k$ is limited to the range $|x|<\ell$.

\section{Using infinite boundary conditions}\label{sec:infinite-system}
\begin{figure}[t]
\includegraphics[width=\columnwidth]{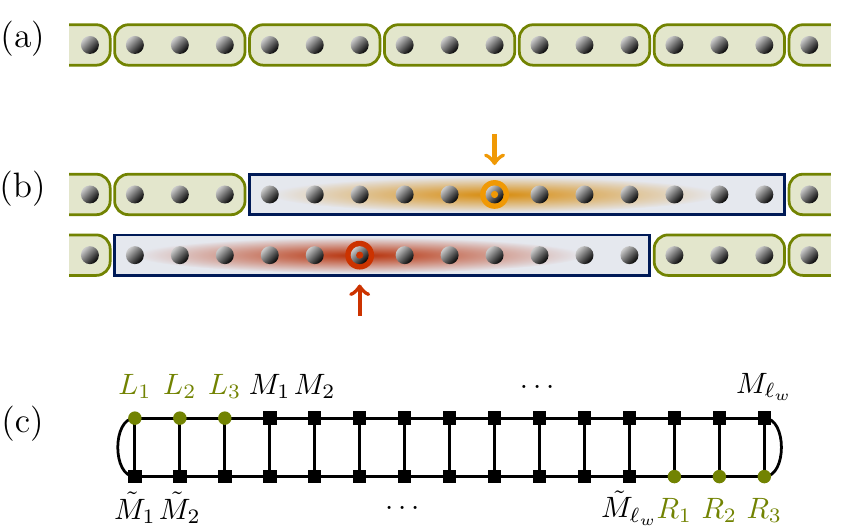}
\caption{\label{fig:iDMRG_scheme}\textbf{Scheme with infinite boundary conditions.} Using infinite boundary conditions for translation-invariant systems, two evolution runs are sufficient to evaluate response functions $S(x,t)$ for a certain spatial range $|x|<\ell$. (a) First, one computes the initial state $\psi$ in uiMPS form. In the diagrams, we assume it has a unit cell of size $n_c=3$. (b) Then, one evolves $\hA^\dag_0|\psi\ket$ and $\hB_0|\psi\ket$ in time by allowing iMPS tensors to vary in a heterogeneous window of width $\ell_w$ around site $0$. (c) Finally, one computes overlaps of the evolved states shifted spatially relative to each other to obtain $S(x,t)$.}
\end{figure}
\subsection{The method}
An attractive option for the evaluation of response functions $S(x,t)=\bra\psi|\hA_x(t)\hB_0(0)|\psi\ket$ for a whole range of distances $x$ is to work in the thermodynamic limit using infinite boundary conditions. In this case, one can use the following procedure, for which two time-evolution runs are sufficient.
\begin{enumerate}
 \item
 Compute a uniform iMPS (uiMPS) approximation $\psi$ of the ground state using imaginary-time evolution \cite{Vidal2007-98,Orus2008-78}, iDMRG \cite{White1992-11,White1993-10,McCulloch2008_04}, or the approach of Ref.~\cite{Zauner2018-97} to find iDMRG fixed points with non-local updates.
 \item
 For the following, allow the tensors of the iMPS within a spatial interval (``window'') of appropriate size $\ell_w$ to vary in time as in Refs.~\cite{Phien2012-86,Milsted2013-88}, i.e., switch from a uniform to a heterogeneous iMPS.
 \item
 Apply operators $\hA$ and $\hB$ at a site $i=0$ in the center of the window to get $\hA^\dag_0|\psi\ket$ and $\hB_0|\psi\ket$.
 \item
 Evolve both states in time to obtain $|\psi_{\hA^\dag_0}(-t_1)\ket$ and $|\psi_{\hB_0}(t_2)\ket$.
 \item
 Evaluate overlaps of the time-evolved states with a relative spatial shift by $x$ sites to obtain
 \begin{equation}\label{eq:SxtInfinite}
  	S(x,t_1+t_2)=e^{iE_0(t_1+t_2)}\bra\psi_{\hA^\dag_0}(-t_1)|\hat{T}_{-x}|\psi_{\hB_0}(t_2)\ket,
 \end{equation}
 where $\hat{T}_{-x}$ shifts by $-x$ sites.
\end{enumerate}

Note that, if the system is critical, we are formally working in the thermodynamic limit here, but the finite MPS bond dimension $D$ of the initial state inevitably imposes a finite correlation length. A scaling analysis in $D$ is required to really capture the thermodynamic limit.

While we work with windows of fixed size, it is also possible to reduce computation costs somewhat by starting from a small window around site $0$ and expand it during the time evolution in accordance with the spreading of correlations \cite{Phien2013-88}. Usually, the resulting gains should, however, be minor. As the entanglement growth is limited to causal cones around the perturbations, the computation costs are in any case dominated by the sites inside these cones.

For the evolution of the heterogeneous iMPS, we use a fourth-order Trotter-Suzuki decomposition \cite{Trotter1959,Suzuki1976} of the time-evolution operator. In order to keep the MPS tensors to the left (right) of the window invariant, all Hamiltonian terms outside the window are projected onto the reduced $D$-dimensional Hilbert space of the left (right) block. Let $(i-1,i)$ be the bond at the left boundary of the window. An evolution operator for Hamiltonian terms that are entirely supported in the left block then only act on the left index of the MPS tensor at the left end of the window $[M_i^{\s_i}]_{a,b}\mapsto \sum_{a'} U_{a,a'}[M_i^{\s_i}]_{a',b}$. For the MPS tensor $[M_i^{\s_i}]_{a,b}$ of site $i$, the index $\s_i$ labels local orthonormal basis states and $a,b=1,\dots,D$ label basis states for reduced Hilbert spaces of the left block (sites left of $i$) and the right block (sites right of $i$), respectively. A Hamiltonian term acting on site $i$ and sites in the left block results in a unitary acting on the physical index and left index of that MPS tensor, i.e., $[M_i^{\s_i}]_{a,b}\mapsto \sum_{\s'_i,a'} U^{\s_i,\s_i'}_{a,a'}[M_i^{\s'_i}]_{a',b}$.

The overlap \eqref{eq:SxtInfinite} for a given window size $\ell_w$ and spatial translation $x$ can be evaluated efficiently with a cost that is proportional to $\ell_w+x$. Let $L$ and $R$ be the left and right-orthonormalized variants of the uiMPS tensors of the initial state $\psi$, i.e., $\sum_\s (L^\s)^\dag L^\s=\id$ and $\sum_\s R^\s (R^\s)^\dag=\id$. Then, the left eigen-matrix of the transfer operator $\sum_\s (L^\s)^*\otimes L^\s$ and the right eigen-matrix of the transfer operator $\sum_\s (R^\s)^*\otimes R^\s$ with the (maximum) eigenvalue 1 are $D$-dimensional identity matrices. The overlaps \eqref{eq:SxtInfinite} are obtained by contracting these identity matrices with tensors from the heterogeneous windows of $|\psi_{\hA^\dag_0}(-t_1)\ket$ and $|\psi_{\hB_0}(t_2)\ket$ and $x$ tensors from $\psi$ at each end as shown in Fig.~\ref{fig:iDMRG_scheme}.

In all simulations, we use a Trotter time step $\Delta t=0.1$ such that truncation errors dominate. Precision and computation costs are controlled by truncating state components with Schmidt coefficients $\lambda_k<\lambda_\trunc$ below a suitably chosen truncation threshold $\lambda_\trunc$ or by fixing the bond dimension $D$. These parameters are specified in the figure captions.
\begin{figure}[t]
\includegraphics[width=\columnwidth]{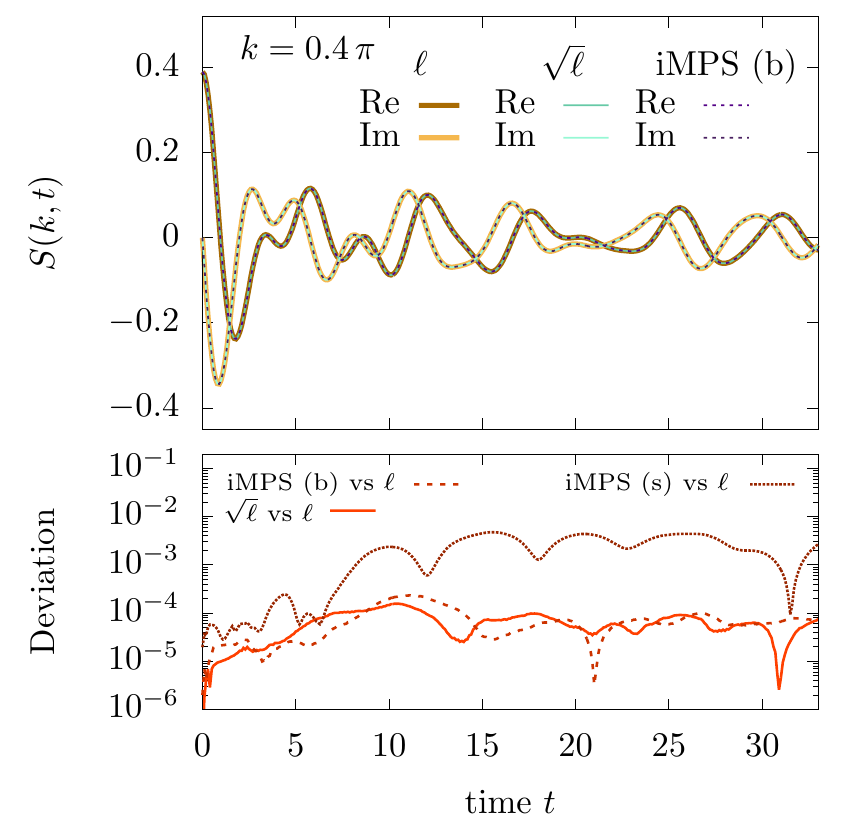}
\caption{\label{fig:compare}\textbf{Comparison of different schemes.} Response functions $S(k,t)=\sum_{x}e^{-ikx}\bra\psi|\hS^\alpha_x(t)\hS^\alpha_0(0)|\psi\ket$ computed with the different MPS schemes are compared for the spin-1 system \eqref{eq:Hblbq} at $\theta=\pi/5$. Data for the finite-size schemes are labeled by ``$\ell$'' for the original scheme and ``$\sqrt{\ell}$'' for the square-grid scheme, respectively, and we used chains of length $L=256$. For the scheme with infinite boundary conditions, one iDMRG simulation was initialized with a symmetric state [iMPS (s)]; another simulation was initialized with a state that breaks the spin-flip symmetry [iMPS (b)] and features, for the same $\lambda_\trunc$, smaller bond dimensions. Truncation thresholds were chosen to be $\lambda_\trunc^2 = 10^{-9}$ for the ground-state computation and $10^{-8}$ during the time evolution.}
\end{figure}

\subsection{Comparison of the schemes}
\begin{figure}[t]
\includegraphics[width=\columnwidth]{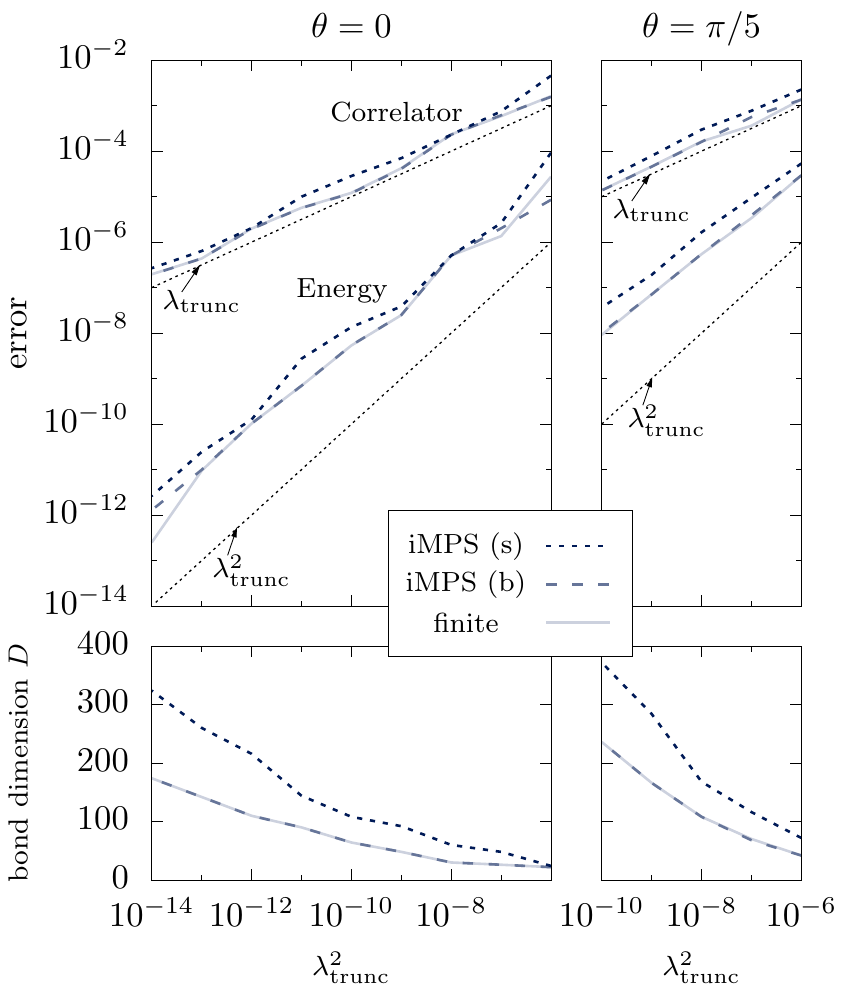}
\caption{\label{fig:compare_gs}\textbf{iDMRG with symmetric and symmetry-broken initial states.} As a function of the truncation threshold $\lambda^2_\trunc$, we compare the results of iDMRG simulations with symmetric [iMPS (s)] and symmetry-broken [iMPS (b)] initial states and finite-size ground-state simulations ($L=256$) for the bilinear-biquadratic spin-1 chain at $\theta=0$ (left) and $\pi/5$ (right). The top panels show the maximum error of the correlator, $\max_i \vert\bra\hS^z_i\hS^z_0\ket_{\lambda_\trunc} - \bra\hS^z_i\hS^z_0\ket_\text{exact}\vert$ and the error of the ground-state energy per site, both compared to quasi-exact reference values. The dashed black lines indicate $\lambda_\trunc$ and $\lambda^2_\trunc$ as a reference. In the bottom panels, we compare the bond dimensions in these simulations.}
\end{figure}

Figure~\ref{fig:compare} compares a simulation with infinite boundary conditions to simulations using the two finite-size schemes. The iMPS window size $\ell_w$ is chosen such that the perturbation, spreading from the center, does not reach the window boundaries for the considered times. Deviations between the different schemes are not discernible in the top panel. Deviations between the two finite-size schemes are small. They are due to the fact that different sets of lattice sites are used and are consistent with the chosen truncation threshold, i.e., of order $\mc{O}(\lambda_\trunc)$. We observe differences between two iMPS simulations that only differ in their initial states. Data labeled ``iMPS (s)'' were computed using a symmetric initial state for the iDMRG; the exact 2-site ground state. Data labeled ``iMPS (b)'' were computed by starting from the symmetry-broken state $\left|\uparrow,\downarrow\right\ket$. The deviations of the symmetry-broken simulation are similar to those of the finite-size schemes (also initialized with symmetry-broken states); the deviations of the symmetric simulation are larger. To clarify this issue, ground-state energies, correlation functions, and bond dimensions are compared in Fig.~\ref{fig:compare_gs}. For the same truncation threshold, errors and bond dimensions of the symmetry-broken iDMRG simulations are very similar to those of simulations for finite but long chains ($L=256$). The symmetric iDMRG simulations yield, as a function of the truncation threshold, slightly larger errors and roughly doubled bond dimensions. This can be explained as follows.

For symmetry-broken initial states, the iDMRG converges to one of two fixed points, corresponding to two uiMPS $|\psi^{(b)}\ket$ and $\hF|\psi^{(b)}\ket$, where $\hF=\bigotimes_i\hF_i$ denotes the global spin-flip operator. Block Hilbert spaces for these states are orthogonal. 
For symmetric initial states, the iDMRG converges to an equal-weight superposition $|\psi^{(s)}\ket=(|\psi^{(b)}\ket+\hF|\psi^{(b)}\ket)/\sqrt{2}$. The two symmetry-broken fixed points have the same Schmidt spectrum $\{\lambda_k\}$. Due to the orthogonality of the block Hilbert spaces, for each $\lambda_k$, the Schmidt coefficient $\lambda_k/\sqrt{2}$ occurs twice in the Schmidt decomposition of $|\psi^{(s)}\ket$. Hence, the symmetric simulation requires roughly twice the bond dimension of $|\psi^{(b)}\ket$ for a given truncation threshold. In particular, a symmetric computation with lowered truncation threshold $\lambda_\trunc/\sqrt{2}$ requires bond dimension $2D_b$ where $D_b$ is the bond dimension of the symmetry-broken simulation with truncation threshold $\lambda_\trunc$.

\subsection{Degeneracy of the transfer operator}
There is another good reason for avoiding the symmetric initial states in iDMRG simulations: The orthogonalization of the resulting uiMPS $|\psi\ket$ \cite{Orus2008-78} requires the computation of dominant eigen-matrices of MPS transfer operators. For spin-flip symmetric states, eigenvalues of the transfer operators are doubly degenerate. If the spin-flip symmetry is not imposed explicitly, the degeneracy can be slightly broken. If this is not handled with care, the orthogonalization of the wavefunction will be imprecise, impairing the accuracy of observables.

For a bipartition of the system, let the Schmidt decomposition of the MPS be denoted by $|\psi\ket=\sum_k\lambda_k|\ell_k\ket\otimes |r_k\ket$. For a spin-flip symmetric state $\hF|\psi\ket=f|\psi\ket$ with $f=\pm 1$, each Schmidt component is either spin-flip symmetric ($\hF|\ell_k\ket\otimes|r_k\ket=f|\ell_k\ket\otimes|r_k\ket$) or there exists a second component $k'$ with $\lambda_{k'}=\lambda_{k}$ and $\hF|\ell_k\ket\otimes|r_k\ket=f|\ell_k'\ket\otimes|r_k'\ket$. Hence, after appropriate unitary transformations, all block basis states are spin-flip symmetric with the product of their eigenvalues $\pm 1$ being equal to $f$, and MPS tensors of the spin-flip symmetric state obey the condition
\begin{equation}\label{eq:flipSymmetry}
	\sum_{\sigma'}\bra\sigma|\hF_i|\sigma'\ket \mc{F} A^{\sigma'}\mc{F}^\dag=A^{\sigma}
\end{equation}
with spin-flip operators on the bond vector spaces denoted by $\mc{F}$.
Specifically, in a uiMPS defined by tensors $A$ for an entire unit cell, the left and right bond vector spaces and corresponding flip operators $\mc{F}$ in Eq.~\eqref{eq:flipSymmetry} coincide. Let $V$ be a dominant eigen-matrix of the MPS transfer operator, i.e., $\sum_\sigma(A^\sigma)^\dag VA^\sigma=\eta V$. Then, it follows from Eq.~\eqref{eq:flipSymmetry} that $\mc{F}^\dag V\mc{F}$ is also an eigen-matrix with the same eigenvalue $\eta$, explaining the degeneracy of the transfer operator.

\section{Results for bilinear-biquadratic spin-1 chains}\label{sec:results}
\begin{figure}[t]
\includegraphics[width=\columnwidth]{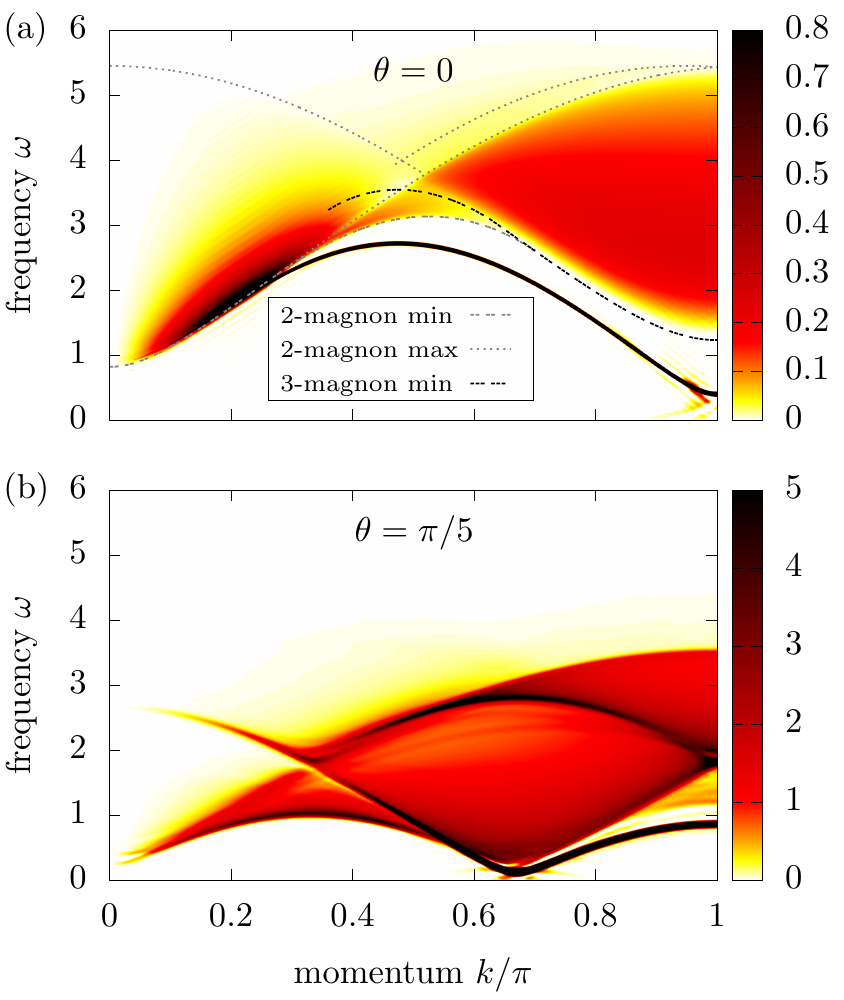}
\caption{\label{fig:spin1-S}\textbf{Dynamic structure factor in spin-1 chains.} Numerical results for the dynamic spin structure factor \eqref{eq:S} in the bilinear-biquadratic spin-1 chain \eqref{eq:Hblbq} at $\theta=0$ and $\theta=\pi/5$ in the Haldane phase. The MPS simulations were done with infinite boundary conditions and, during the time evolution, truncation thresholds were set to $\lambda^2_\trunc = 10^{-10}$ and $\lambda^2_\trunc=10^{-8}$ for $\theta=0$ and $\theta=\pi/5$, respectively. The real-time response functions have been extrapolated by linear prediction before the Fourier transform.
}
\end{figure}

Figure~\ref{fig:spin1-S} shows numerical results for the dynamic spin structure factor
\begin{equation}\label{eq:S}
	S(k,\omega)=\sum_{x}e^{-ikx}\int\ud t\, e^{i\omega t}\bra\psi|\hS^\alpha_x(t)\hS^\alpha_0(0)|\psi\ket
\end{equation}
in the spin-1 Heisenberg antiferromagnet ($\theta=0$) and the bilinear-biquadratic chain at $\theta=\pi/5$ which is still in the Haldane phase, but close to the critical Uimin-Lai-Sutherland (ULS) point at $\theta=\pi/4$, where the model is Bethe-ansatz integrable \cite{Uimin1970-12r,Uimin1970-12,Lai1974-15,Sutherland1975-12}. Note that, due to the rotation invariance of the model, $S(k,\omega)$ is the same for all projections of the spin components $\alpha=x,y,z$ or, in fact, any projection of the spin operators.
The spectral weight $\int\ud\omega\, S(k,\omega)=2\pi\bra\psi|\hS^\alpha_k\hS^\alpha_0|\psi\ket$ (static spin structure factor) vanishes for $k\to 0$ and all $\theta$, as the ground states have zero total $\hS_{k=0}^z$ quantum number.

Let us first discuss the Heisenberg antiferromagnet at $\theta=0$. Around $k\approx 0$ and $k\approx\pi$, the low-energy physics is approximately captured by the $O(3)$ nonlinear $\sigma$ model (NL$\sigma$M) which is based on a large-$s$ expansion \cite{Haldane1983-93,Affleck1989-1,Fradkin2013}. For integer spin chains, this field theory equivalently describes a classical 2d ferromagnet at an effective finite temperature. Hence, correlations decay exponentially, which led Haldane to conjecture a finite excitation energy gap for such antiferromagnetic integer spin chains \cite{Haldane1983-50,Haldane1983-93}.

\begin{figure}[t]
\includegraphics[width=0.94\columnwidth]{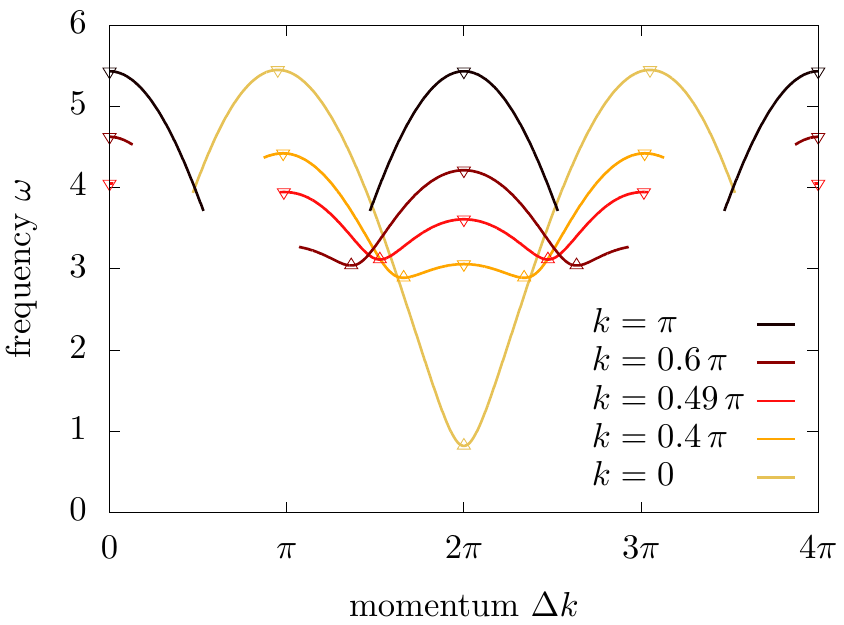}
\caption{\label{fig:spin1-HAFM-2magnon}\textbf{Two-magnon dispersion relation for $\theta=0$.} For the Heisenberg antiferromagnet with $\theta=0$ in Eq.~\eqref{eq:Hblbq}, the two-magnon dispersion relation is shown for several $k=k_1+k_2$ as a function of $\Delta k=k_1-k_2$. End points result from the fact that the single-magnon excitation enters the two-magnon continuum at $k_i\approx 0.24\pi$. The global maxima and minima define the boundaries of the two-magnon continuum and the submaxima yield sudden jumps in the two-magnon density of states. The corresponding boundaries are indicated in Fig.~\ref{fig:spin1-S}(a).}
\end{figure}

The Haldane gap is clearly visible in Fig.~\ref{fig:spin1-S}(a) at $k=\pi$ and has a value of $\Delta E_\text{H}\approx 0.41048$ \cite{Nightingale1986-33,White2008-77}. The lowest excited states are a triplet of spin-1 magnons. These single-particle excitations result in a delta peak in the dynamic structure factor. The additional small low-energy features in the structure factor around $k=\pi$ are associated with the edge states \cite{Affleck1988-115,Kennedy1990-2,Pollmann2012-85} of the model.
The single-magnon energy increases as $k$ is lowered, goes through an inflection point and a maximum, before entering the two-magnon continuum at $k\approx 0.24\,\pi$ \cite{Takahashi1994-50,Yamamoto1997-235,White2008-77}.
Neglecting magnon-magnon interactions, upper and lower thresholds for the two-magnon continuum can be obtained from the single-magnon dispersion. In addition, due to a maximum in the two-magnon dispersion relation, a van Hove singularity occurs below the upper threshold of the continuum. At this point, the two-magnon density of states drops discontinuously. All of these thresholds are indicated in Fig.~\ref{fig:spin1-S}(a) and can be understood by inspecting the two-magnon dispersions shown in Fig.~\ref{fig:spin1-HAFM-2magnon}. The two-magnon continuum has its minimum of $\approx 2\Delta E_\text{H}$ at $k=0$. The structure factor is considerably increased in the region where the single-magnon branch enters. While no strong features appear at the upper boundary of the continuum, the van Hove singularity results in a clear sudden drop of the structure factor.

Similarly, the three-magnon continuum has its minimum of $\approx 3\Delta E_\text{H}$ at $k=\pi$ and we again indicate the approximate lower threshold. This and the two-magnon thresholds explain to a large extent the non-trivial structure of $S(k,\omega)$ around $(k,\omega)\approx (\pi/2,3.5)$. The shape of the three-magnon contribution to the dynamic structure factor at $k\approx\pi$ has been computed for the integrable NL$\sigma$M \cite{Balog1997-500,Horton1999-60,Essler2000-62}. As discussed in Ref.~\cite{White2008-77}, the peak height and the high-frequency decay, however, deviate quite strongly from the actual structure factor in the Heisenberg antiferromagnet. This is mainly due to a missing UV cut-off in the field-theoretic description.

\begin{figure}[t]
\includegraphics[width=0.94\columnwidth]{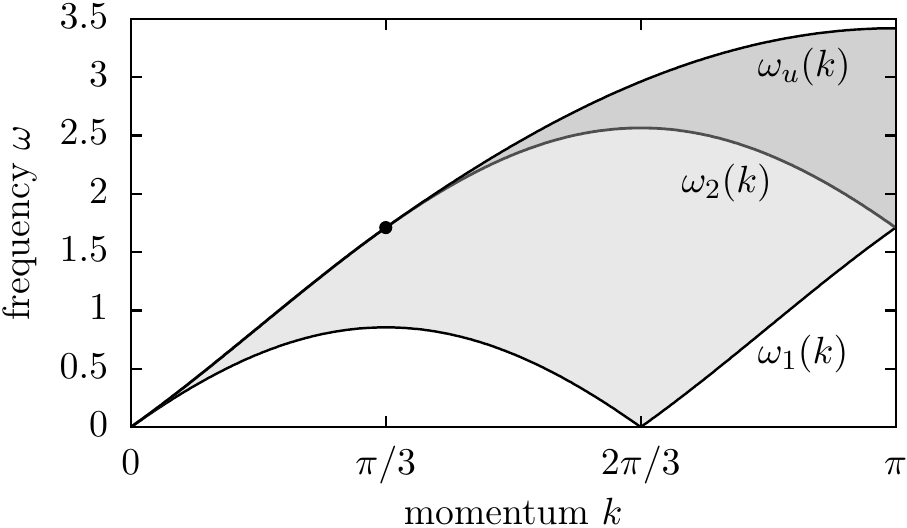}
\caption{\label{fig:spin1-ULS}\textbf{Continuum of low-energy excitations at ULS point.} At the ULS point $\theta=\pi/4$, the spin-1 chain \eqref{eq:Hblbq} can be solved by the nested Bethe ansatz. The figure shows three thresholds for low-energy excitations which consist of two elementary excitations \eqref{eq:BAexcitations}.
The two-particle continuum is bounded from below by $\omega_1(k)$ and from above by $\omega_u(k)$. Below the threshold $\omega_2(k)$, there is one solution for each total momentum $k$. Above $\omega_2(k)$, there are two solutions.}
\end{figure}

When $\theta$ is increased, $S(k,\omega)$ gets restructured considerably, especially, when passing the AKLT point $\theta=\arctan 1/3\approx \pi/10$. At $\theta=\pi/5$ [Fig.~\ref{fig:spin1-S}(b)], the single-magnon mode has undergone drastic changes. It now has a maximum of $\approx 0.87$ at $k=\pi$ and decays almost to zero at $k=2\pi/3$. This and other features can be explained best by comparing to the ULS point $\theta=\pi/4$, where the system has an enlarged $\SU(3)$ symmetry and can be solved by the nested Bethe ansatz \cite{Uimin1970-12r,Uimin1970-12,Lai1974-15,Sutherland1975-12}. The ground state is then parametrized by two sets of rapidities (Bethe quantum numbers). Due to symmetry constraints, the low-energy excitations consist of two soliton-like elementary excitations. They have energies
\begin{subequations}\label{eq:BAexcitations}
\begin{align}
	E(k_1,k_2)&=\veps_1(k_1)+\veps_2(k_2)\quad\text{with}\\
	\veps_1(k_1)=&\frac{\pi\sqrt{2}}{3}\,\frac{\cos(\pi/3-k_1)-\cos(\pi/3)}{\sin(\pi/3)}\quad\text{and}\\
	\veps_2(k_2)=&\frac{\pi\sqrt{2}}{3}\,\frac{\cos(\pi/3)-\cos(\pi/3+k_2)}{\sin(\pi/3)},
\end{align}
\end{subequations}
where $0\leq k_1\leq 2\pi/3$ and $0\leq k_2\leq 4\pi/3$ \cite{Sutherland1975-12,Johannesson1986-270}. The total momentum $k=k_1+k_2\in[0,2\pi)$ can be folded into the first Brillouin zone. The resulting two-particle continuum features three thresholds $\omega_1(k)\leq \omega_2(k)\leq \omega_u(k)$ shown in Fig.~\ref{fig:spin1-ULS}. The lowest energy excitations for given total momentum $k$ have energy $E(k,0)$ for $|k|\leq 2\pi/3$ and $E(2\pi/3,k-2\pi/3)$ for $2\pi/3\leq |k|\leq \pi$. Above the threshold $\omega_2(k)=E(0,k)$, there are two solutions for $(k_1,k_2)$ with given total momentum $k$ and energy instead of just one. Correspondingly, the density of states increases by a factor of 2 at $\omega_2(k)$. The two-particle continuum ends at the upper threshold $\omega_u(k)$ which coincides with $\omega_2(k)$ for $0\leq |k|\leq \pi/3$ and is given by $E(q(k),k-q(k))$ with $q(k)=-\pi/6-\arctan[(\cos(k)-1)/\sin(k)]$ otherwise. For $|k|>\pi/3$, the two-particle density of states has a square-root van Hove singularity at $\omega=\omega_u(k)$ due to the maxima in the two-particle dispersion relations.

These characteristics are reflected in the dynamic structure factor for $\theta=\pi/5$ which is shown in Fig.~\ref{fig:spin1-S}(b). The additional small low-energy features around $k=2\pi/3$ can be due to the edge states or might also be artifacts as low-frequency features are more difficult to resolve with the finite-time data. The splitting of the continuum around $k=0$ indicates that the gap due to breaking of the $\SU(3)$ symmetry is symmetry-sector dependent.
Also, $(n>2)$-particle continua contribute to $S(k,\omega)$. 
For $\theta=\pi/5$, the gap at $k=2\pi/3$ is quite small. This is because the transition from the gapless quadrupolar phase $\pi/2>\theta>\pi/4$ to the Haldane phase $\pi/4>\theta>-\pi/4$ is a Berezinskii-Kosterlitz-Thouless transition \cite{Fath1993-47,Itoi1997-55}. Hence, we have a slow exponential opening of the gap.
We interpret the feature in the region $(k, \omega) \approx (0.2\pi, 2.5)$ as a remnant of the single-magnon excitation. A detailed analysis also at other values of $\theta$ will be presented elsewhere.

When $\theta$ is increased, starting from $\theta=0$, a significant amount of spectral weight is 
transferred from the single-magnon excitation to the multi-magnon continua and the single-magnon band flattens. Around the AKLT point, the single-magnon dispersion becomes monotonic, i.e., ceases to have a maximum around $k=\pi/2$. Increasing $\theta$ further, the dispersion develops a minimum around $k=2\pi/3$ (see also Ref.~\cite{Okunishi2001-64}) and the gap at this point closes for $\theta\to\pi/4$. Correspondingly, when increasing $\theta$, the multi-magnon continua deform and become tighter. At the AKLT point, most of their spectral weight at $k=\pi$ is for example concentrated around $\omega\approx 3.2$. When $\theta$ is increased further, the continua widen again with the lower edge approaching zero at $k=0$ and $k=2\pi/3$ for $\theta\to\pi/4$ and most features corresponding to those of the Bethe ansatz solution at the ULS point $\theta=\pi/4$ (Fig.~\ref{fig:spin1-ULS}).

\section{Cycles in infinite-system DMRG}\label{sec:cycles}
We are using iDMRG \cite{White1992-11,White1993-10,McCulloch2008_04} to compute uiMPS approximations for ground states of bilinear-biquadratic spin-1 chains \eqref{eq:Hblbq} in the thermodynamic limit. For the gapless quadrupolar phase $\pi/4<\theta<\pi/2$, we notice that iDMRG with insertion of 2-site unit cells does not converge in the traditional sense. Interpreted as a discrete dynamical map, it does not converge to 1-cycles but to 3-cycles. While the former correspond to uiMPS with a 2-site unit cell, the latter correspond to uiMPS with a 6-site unit cell. This behavior can be explained by the dominance of quadrupolar period-three correlations in this gapless phase \cite{Fath1991-44,Xian1993-5,Bursill1995-28,Laeuchli2006-74}. As discussed in more detail below, the iDMRG then favors low-entangled states that break the translation invariance with unit cells containing a multiple of 3 sites.

For fermionic systems, convergence of iDMRG to non-trivial limit cycles has been observed earlier on through cycles in the energy density \cite{Caprara1997}. In this case, the phenomenon is attributed to the discrete changes in the number of fermions as the system size is increased iteratively. Also, different from our observations in the spin-1 chains, the variations of energy densities in the cycles were found to be on the order of the truncation error.

\subsection{Summary of the iDMRG algorithm}
The idea of iDMRG, due to Steve White \cite{White1992-11,White1993-10}, is to build up the lattice iteratively and compute, in the process, MPS approximations for the ground states that live in suitably reduced Hilbert spaces. In every iteration, $n_c$ sites are added at the center of the system. Using reduced bases for left and right blocks from the previous iteration, the energy is minimized and bases for the enlarged blocks are selected based on Schmidt decompositions of the ground-state approximation. Operator matrix elements needed for subsequent iterations are projected onto the new bases, in particular, those of the Hamiltonian.

For $n_c=2$, the algorithm can be summarized as follows. After $n$ iterations, the ground-state approximation for the $2n$-site system has MPS form
\begin{equation}\label{eq:MPS}
	|\psi_n\ket = \sum_\vs L^{\s_1}_1 L^{\s_2}_2 \dots L^{\s_n}_n \Lambda_n R^{\bs_n}_n \dots R^{\bs_2}_2 R^{\bs_1}_1 |\vs\ket.
\end{equation}
Tensors $L_i$ and $R_i$ obey left and right orthonormality constraints, respectively. $\Lambda_n$ is a diagonal $D\times D$ matrix containing the Schmidt coefficients of the state $\psi_n$. $D$ is the bond dimension. To obtain a ground-state approximation $\psi_{n+1}$ in the next iDMRG iteration, one replaces $\Lambda_n$ by a 2-site tensor $C^{\s_{n+1}\bs_{n+1}}$, minimizes the energy expectation value with respect to $C$, and applies a singular value decomposition to split $C$ into $L^{\s_{n+1}}_{n+1} \Lambda_{n+1} R^{\bs_{n+1}}_{n+1}$. In that last step, some of the smallest Schmidt coefficients $\lambda_1\geq\lambda_2\geq\dots $ can be discarded to limit the bond dimension $D$ of the resulting MPS $\psi_{n+1}$. The discarded weight
\begin{equation}\label{eq:truncError}\textstyle
	\epsilon:=\sum_{k>D}\lambda^2_k
\end{equation}
is called the truncation error and can be used for convergence analysis.

An addition to the iDMRG algorithm, the wavefunction prediction, was introduced by Ian McCulloch \cite{McCulloch2008_04}. Moving the orthogonality center in the state $\psi_n$ one site to the left or right, one computes
\begin{equation*}
	L^{\s}_n \Lambda_n R^{\s'}_n 
	=:\Lambda^L_n \tilde{R}^{\s}_{n+1} R^{\s'}_n
	=:L^{\s}_n \tilde{L}^{\s'}_{n+1} \Lambda^R_n.
\end{equation*}
This then provides a prediction for $\psi_{n+1}$:
\begin{equation}\label{eq:psiPred}
	|\psi^\pred_{n+1}\ket=\dots L^{\s_n}_n (\tilde{L}^{\s_{n+1}}_{n+1} \Lambda^R_n \Lambda^{-1}_{n-1} \Lambda^L_n \tilde{R}^{\bs_{n+1}}_{n+1} ) R^{\bs_n}_n\dots
\end{equation}
The corresponding prediction for $C$ is used to initialize the energy minimization problem for $\psi_{n+1}$ and reduces the number of required iterations. Furthermore, the prediction infidelity
\begin{equation}\label{eq:Infidel}
	1-|\bra\psi_{n+1}^\pred| \psi_{n+1} \ket|
\end{equation}
can be used to assess the convergence of the algorithm.

Once the algorithm has converged to an acceptable precision (at iteration step $n$), it corresponds to the uiMPS
\begin{equation}\label{eq:uiMPS}
	|\psi_\infty\ket=\dots(L^{\s_i}_n \Lambda_n R^{\s_{i+1}}_n \Lambda^{-1}_{n-1})(L^{\s_{i+2}}_n \Lambda_n R^{\s_{i+3}}_n \Lambda^{-1}_{n-1})\dots
\end{equation}
In a final step, this uiMPS should be orthonormalized which can be done as described in Ref.~\cite{Orus2008-78}.

For iDMRG with cell size $n_c>2$, the algorithm is to be modified as follows. In the wavefunction prediction, the orthogonality center is shifted by $n_c/2$ sites to the right and to the left. One inserts the corresponding $n_c$ tensors at the center of the MPS. For odd $n_c$, the new cell can be inserted alternatingly one site to the left and one site to to right of the center of the previously inserted cell. This way, the left and right blocks are grown equally. For the energy minimization, one sweeps back and forth through the inserted cell and minimizes the energy expectation value, e.g., as in standard 2-site finite-system DMRG. If the algorithm converges to a fixed point (1-cycle), it yields a uiMPS with cell size $n_c$ in analogy to Eq.~\eqref{eq:uiMPS}.

\subsection{Nontrivial limit cycles}
\begin{figure*}[t]
\includegraphics[width=0.9\textwidth]{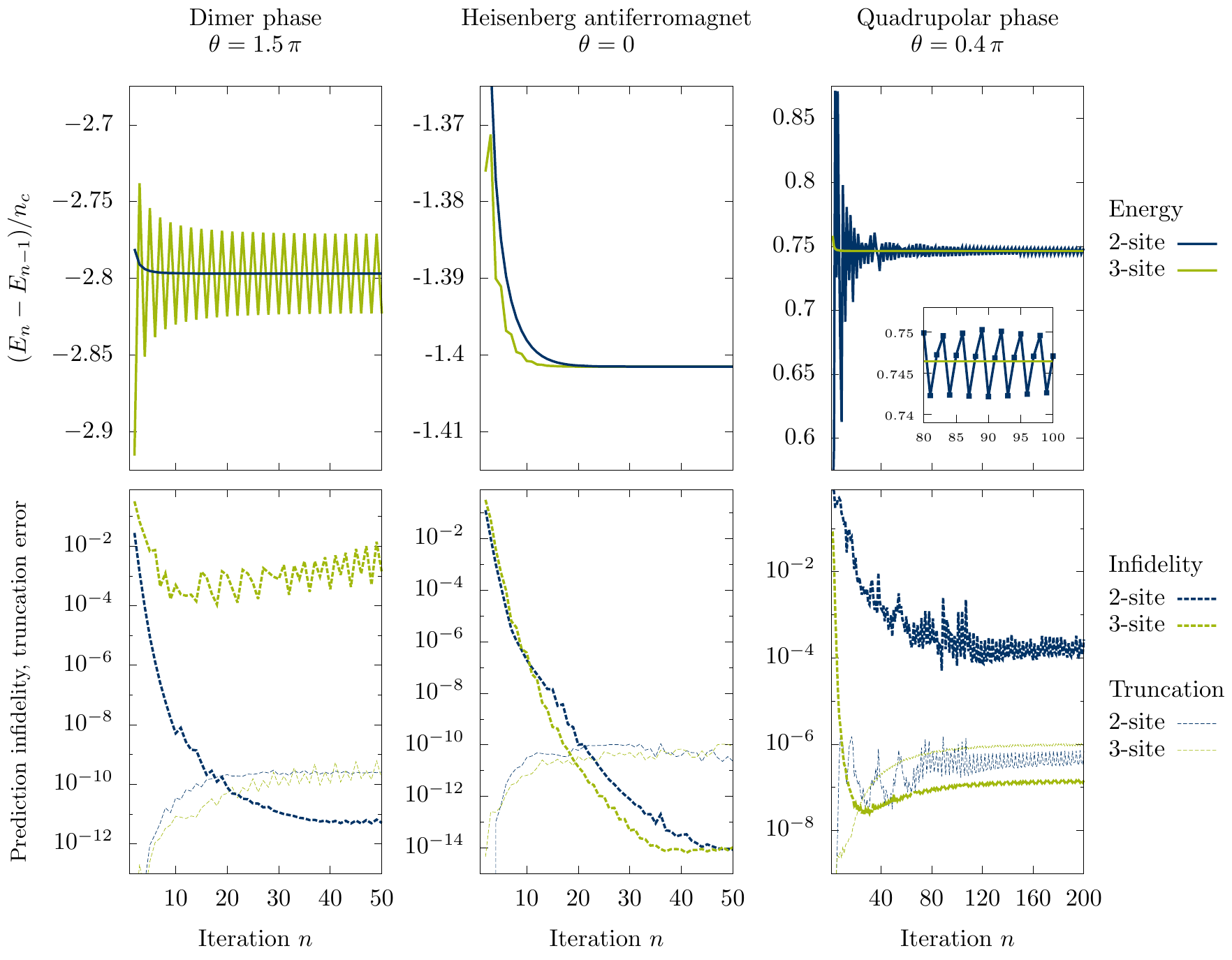}
\caption{\label{fig:iDMRGconvergence}\textbf{Convergence of iDMRG.} The columns refer to the dimer phase (left, $\theta = 1.5\,\pi$), the Heisenberg antiferromagnet (center, $\theta = 0$), and the gapless quadrupolar phase (right, $\theta = 0.4\,\pi$). As a function of the iteration $n$, the top panels show the change $(E_n - E_{n-1})/n_c$ of the total energy due to the insertion of one unit cell, divided by the cell size $n_c$. The bottom panels display the prediction infidelity $1-|\bra\psi_{n}^\pred| \psi_{n} \ket|$ (thick lines) and the truncation error $\epsilon$ (thin lines). The line colors indicate the unit-cell size used in the simulations -- either $n_c=2$ or $n_c=3$ sites. Bond dimensions for all simulations in this plot were set to $D=400$.}
\end{figure*}
Figure~\ref{fig:iDMRGconvergence} shows convergence properties of iDMRG with 2-site and 3-site unit cells in different phases of the bilinear-biquadratic spin-1 chain \eqref{eq:Hblbq}. For the Heisenberg antiferromagnet at $\theta=0$, the algorithm converges well for both cell sizes. At $\theta=1.5\,\pi$ in the dimerized phase, iDMRG with cell size $n_c=2$ converges well, but for $n_c=3$ we see period-2 oscillations. As we will discuss in more detail for similar observations in the quadrupolar phase, this does not mean that $n_c=3$ iDMRG has not converged. Rather, it converges to a limit $2$-cycle that corresponds to a uiMPS with cell size 6. As a multiple of two, this is commensurate with a spontaneous dimerization of the ground state.

\begin{figure}[t]
\includegraphics[width=0.9\columnwidth]{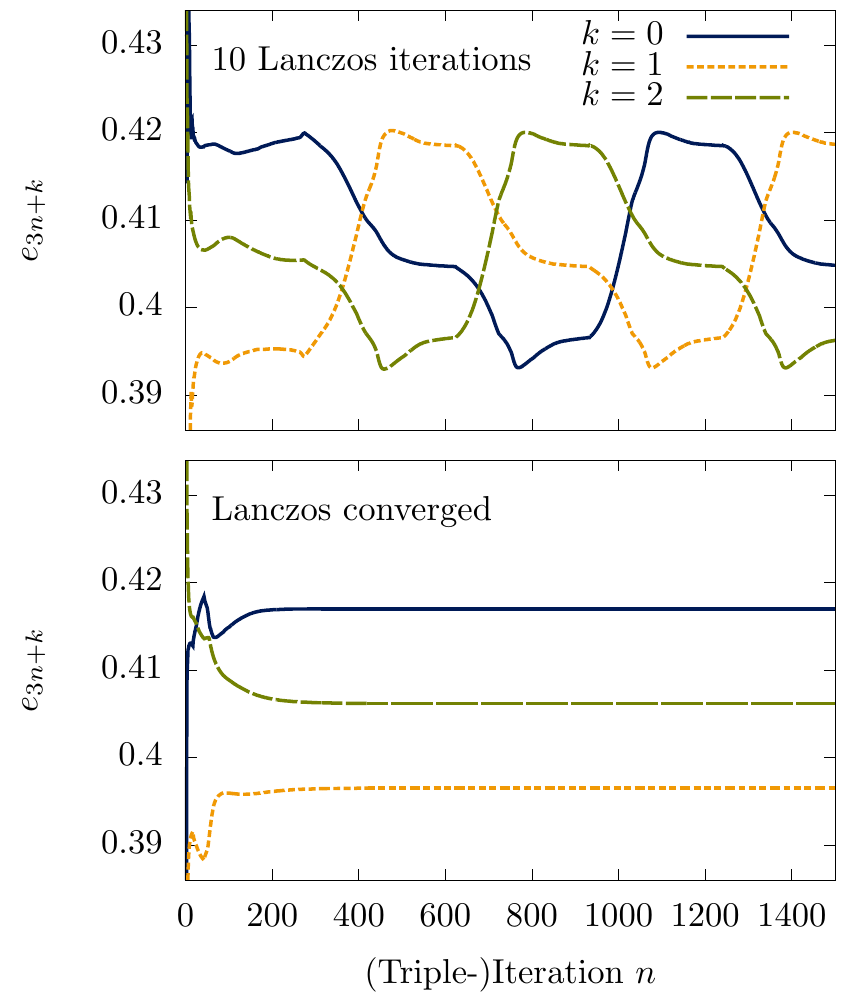}
\caption{\label{fig:n-cycles}\textbf{Nontrivial limit cycles in iDMRG.} For a simulation in the gapless quadrupolar phase of the spin-1 chain \eqref{eq:Hblbq} at $\theta = 0.3\,\pi$, we assess the convergence of iDMRG with a $2$-site unit cell. The plots show the change $e_{3n+k}=(E_{3n+k}-E_{3n+k-1})/n_c$ in the total energy due to the insertion of a unit cell, divided by the cell size $n_c$, where $k=0, 1, 2$. For the simulation shown in the top panel, $10$ Lanczos iterations were done for the energy minimization in each iteration. For the simulation shown in the bottom panel, the local energy minimization was run until convergence. This stabilizes a limit cycle of length 3 and then corresponds to a uiMPS with a 6-site unit cell. Here, the bond dimension was set to $D=100$.}
\end{figure}

\begin{figure}[!t]
\includegraphics[width=\columnwidth]{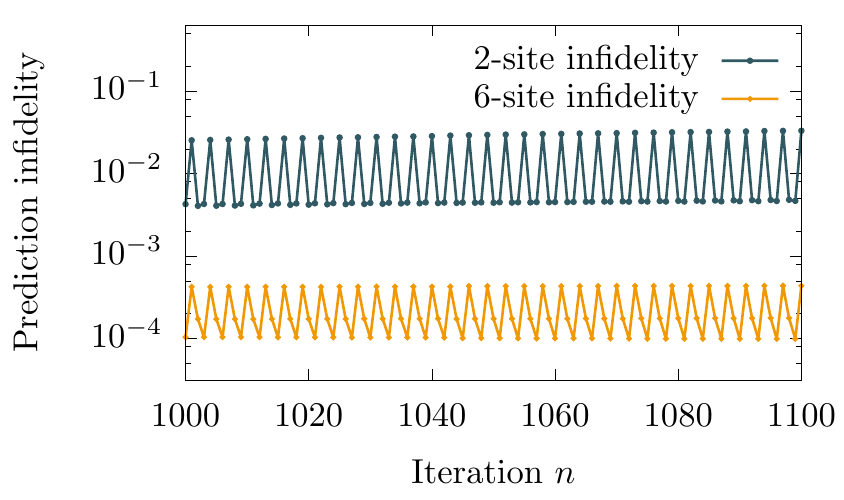}
\caption{\label{fig:n-cycles-fidel}\textbf{Prediction fidelity for nontrivial cycles.} Here, we consider the same 2-site cell iDMRG simulation as in the lower panel of Fig.~\ref{fig:n-cycles}. We show the regular (2-site) infidelity \eqref{eq:Infidel} which compares the state after one iteration to the state obtained by insertion of a $2$-site cell using the wavefunction prediction. Secondly, the 6-site infidelity compares the state after three iterations to the state obtained by insertion of a $6$-site cell using the wavefunction prediction. The quantities are shown after a large number of iterations and the Lanczos algorithm for energy minimizations is run until convergence.}
\end{figure}

In the critical quadrupolar phase, both energy and infidelity indicate very good convergence for $n_c=3$, but there are strong oscillations and a rather high prediction infidelity for $n_c=2$. As shown in an inset, the oscillations have a period of three but are not really stable. Figure~\ref{fig:n-cycles} shows that the oscillations become stable 3-cycles when the number of Lanczos iterations in the energy minimizations is increased until convergence. This can be explained as follows. The critical phase features dominant period-three correlations \cite{Fath1991-44,Xian1993-5,Bursill1995-28,Laeuchli2006-74}. In simulations with $n_c=3$, iDMRG can converge to a ground-state approximation with translation invariance spontaneously broken to an invariance under shifts by three sites. These symmetry broken states $\psi^{(3)}$ have lower entanglement than a corresponding translation-invariant state $|\psi^*\ket=|\psi^{(3)}\ket+\hat{T}_{1}|\psi^{(3)}\ket+\hat{T}_{2}|\psi^{(3)}\ket$ and are hence favored by iDMRG for a given fixed bond dimension $D$. If iDMRG with $n_c=2$ converged to a fixed point (1-cycle), it would necessarily produce the higher entangled state $\psi^*$. The convergence to the limit 3-cycles means that it rather breaks the translation invariance and yields uiMPS with cell size 6. Of course, the 2-site wavefunction prediction is incorrect in this case, explaining the large 2-site infidelities in Figs.~\ref{fig:iDMRGconvergence} and \ref{fig:n-cycles-fidel}. That the algorithm with cell size $n_c=2$ has, however, indeed converged to a uiMPS with 6-site unit cell is confirmed by the small 6-site infidelities shown in Fig.~\ref{fig:n-cycles-fidel}. For these, we compare the state $\psi_{n+3}$ -- the state obtained from $\psi_n$ by 3 iterations of $n_c=2$ iDMRG -- to the state $\psi^{\pred,6}_{n+3}$ which is obtained from $\psi_n$ by a 6-site wavefunction prediction without energy optimization. For this, the orthogonality center of $\psi_n$ is rotated three sites to the left and to the right and the resulting 6 tensors are inserted in analogy to Eq.~\eqref{eq:psiPred}.

In conclusion, when iDMRG with cell size $n_c$ converges to a limit cycle of length $m$, the algorithm has converged to a uiMPS with a unit cell of size $mn_c$ instead of $n_c$. This happens naturally when there exist low entangled ground-state approximations with spontaneously broken translation symmetry. In such cases, where iDMRG converges to a nontrivial limit cycle or shows periodic oscillations, it is advisable to change the cell size $n_c$ appropriately to harness the efficiency gains due to the $n_c$-site wavefunction prediction.

\section{Discussion}\label{sec:discussion}
We have demonstrated how the number of time-evolution runs in the computation of spectral functions and dynamic structure factors for translation-invariant systems can be significantly reduced. For finite-system simulations, one can apply the square-grid scheme which reduces the required number of runs from $\ell$ to $2\sqrt{\ell}$, or one can employ momentum-space operators encoded as MPOs if one is only interested in the dynamic response for a few momenta. Alternatively, one can approximate the ground state in form of an iMPS and simulate the time evolution using a finite heterogeneous window with infinite boundary conditions. This version only needs two time-evolution runs, as we can spatially shift the wavefunctions relative to each other to evaluate the response function for all distances. The results of the improved schemes are in very good agreement with simulations using the standard scheme. Hence, they are an attractive technical advancement that allows to significantly reduce the required computation cost. We have applied the technique to compute high-resolution dynamic structure factors for ground states of bilinear-biquadratic spin-1 chains. The results allow us to discuss the low-lying excitations for two points in the Haldane phase with quite different physics due to the influence of the biquadratic interaction. Note that after the central parts of this work were completed, Ref.~\cite{Lange2018-97} appeared. There, infinite boundary conditions are used to compute spin structure factors at finite temperatures.
For the example of the spin-1 chains, we also demonstrated that iDMRG may converge to non-trivial limit cycles. These then correspond to uiMPS with enlarged unit cells, indicating the existence of low-entangled states with reduced translation invariance. We also showed that initializing iDMRG with symmetric states can result in substantially increased bond dimensions and (problematic) degeneracies of the MPS transfer operator. More detailed results on the bilinear-biquadratic spin-1 chains will be presented elsewhere.

\end{document}